 \documentclass[final,5p,times,twocolumn,authoryear]{elsarticle}

\usepackage{amssymb}
\usepackage{amsmath}
\usepackage{hyperref}
\usepackage{lipsum}

\journal{Astronomy $\&$ Computing}

\begin{document}

\begin{frontmatter}



\title{RatanSunPy: A Robust Preprocessing Pipeline for RATAN-600 Solar Radio Observations Data}



\author[first,forth]{Irina Knyazeva}
\ead{iknyazeva@gmail.com}
\author[first,second]{Igor Lysov}
\ead{ilysov@hse.ru}
\author[first,third]{Evgenii Kurochkin}
\ead{79046155404@yandex.ru}
\author[first,third]{Andrey Shendrik}
\author[second]{Denis Derkach}
\author[first]{Nikolay Makarenko}

\affiliation[first]{organization={Central (Pulkovo) Astronomical Observatory at RAS },
            addressline={Pulkovskoye Shosse 65/1},
            postcode={196140},
            city={Saint-Petersburg},
            country={Russia}}
\affiliation[second]{organization={HSE University},
                    addressline={Myasnitskaya Street 20},
                    postcode={101000},
                    city={Moscow},
                    country={Russia}}
\affiliation[third]{organization={Saint-Petersburg Branch of Special Astrophysical Observatory at RAS},
                    addressline={Pulkovskoye Shosse 65/1, AY},
                    postcode={196140},
                    city={Saint-Petersburg},
                    country={Russia}}
\affiliation[forth]{organization={N. P. Bechtereva Institute of the Human Brain of RAS},
                    addressline={Akademika Pavlova 12A},
                    postcode={197022},
                    city={Saint-Petersburg},
                    country={Russia}}

\begin{abstract}
 The advancement of observational technologies and software for processing and visualizing spectro-polarimetric microwave data obtained with the RATAN-600 radio telescope opens new opportunities for studying the physical characteristics of solar plasma at the levels of the chromosphere and corona. These levels remain some difficult to detect in the ultraviolet and X-ray ranges. The development of such methods allows for more precise investigation of the fine structure and dynamics of the solar atmosphere, thereby deepening our understanding of the processes occurring in these layers. The obtained data also can be utilized for diagnosing solar plasma and forecasting solar activity. However, using RATAN-600 data requires extensive data processing and familiarity with the RATAN-600. This paper introduces \texttt{RatanSunPy}, an open-source Python package developed for accessing, visualizing, and analyzing multi-band radio observations of the Sun from the RATAN-600 solar complex. The package offers comprehensive data processing functionalities, including direct access to raw data, essential processing steps such as calibration and quiet Sun normalization, and tools for analyzing solar activity. This includes automatic detection of local sources, identifying them with NOAA (National Oceanic and Atmospheric Administration) active regions, and further determining parameters for local sources and active regions.  By streamlining data processing workflows, \texttt{RatanSunPy} enables researchers to investigate the fine structure and dynamics of the solar atmosphere more efficiently, contributing to advancements in solar physics and space weather forecasting.

\end{abstract}




\end{frontmatter}




\section{Introduction}
\label{introduction}

Advancements in observational astronomy have ushered in an era characterized by the proliferation of big data, particularly in solar physics. The integration of diverse data types for the same astronomical object, from multiple atmospheric layers and instruments, has become a standard practice. This interdisciplinary approach offers unprecedented opportunities to enhance our understanding of the complex dynamics governing the Sun and its interactions with the heliosphere.

Integrating radio methods and data into the observation and modeling of the heliosphere is critical for advancing our comprehension of the physics behind the heliophysical system and its interconnected components~\citep{bogod2017study}. This importance has been consistently highlighted in numerous studies, such as ~\cite{bisi2015preface}, which underscore the pressing need for advanced tools and frameworks to effectively manage, analyze, and interpret the diverse datasets involved.

In this context, the analysis and interpretation of radio data from the Sun are particularly crucial for gaining deeper insights into solar physics processes. Moreover, radio data is indispensable for the development of effective space weather monitoring and prediction systems. With solar phenomena driving space weather, the ability to analyze radio data can significantly improve our forecasting abilities, providing vital complementary information that spans across all layers of the solar atmosphere~\citep{bogod2018method}. The implications for satellite safety, telecommunications, and power grids on Earth make this work even more essential.

Additionally, radio observations hold the potential to increase our understanding of space weather~\citep{smol'kov2008radiofeatures}, not just in Earth's vicinity but also for other solar system bodies. The detailed study of the Sun-Earth connection, encompassing everything from the solar dynamo to ground-level events, is of utmost importance in a world increasingly dependent on satellite and space-based technologies.

The ability to predict the flare productivity of an active region (AR) during its early emergence is particularly significant~\citep{kutsenko2021possibility}. From an observational standpoint, radio astronomical data can detect pre-flare activity earlier than optical instruments, as demonstrated in several pivotal studies ~\citep{bogod2018method, borovik2012local, abramov2015dynamics, peterova2021increased}.

Instruments like RATAN-600, with their high sensitivity \citep{bogod2011system,khaikin2019radioheliographs} and acceptable spatial resolution \citep{tokhchukova2014computation}, play an essential role in detecting both powerful and faint microwave sources. This ability is crucial for analyzing weak events and studying active regions during their formative stages \citep{grigor2014weak, peterova2021increased}. However, access to and processing of this data has traditionally been a complex task, creating a barrier for researchers eager to explore these rich datasets.

The RATAN-600 instrument boasts a rich history of solar observations, dating back to at least 1997 \citep{tokhchukova2011ratan}, with a comprehensive archive spanning solar cycles 23, 24, and the ongoing 25th cycle. This archive contains more than 50000 observations, representing an invaluable long-term dataset for solar physicists. The data, provided as 1D scans in both left and right circular polarizations, covers a frequency range of 3–18 GHz~\citep{shendrik2020spatial}. Such an extensive archive offers unique opportunities for longitudinal studies of solar phenomena, yet without efficient data processing tools, its full potential remains largely untapped.

Given the volume and complexity of data from solar radio observations, there is an urgent need for modern, accessible tools to facilitate data processing and analysis. Tools like the \href{https://sunpy.org/}{SunPy library} have made solar data more accessible, allowing researchers from diverse fields, including computer science, to contribute to solar physics. This development has led to a substantial increase in research outputs, further underlining the importance of developing specialized Python packages tailored for radio data processing. Such packages can democratize access to crucial datasets, accelerating discoveries and fostering interdisciplinary collaborations across the academic community.

Unfortunately, the radio-astronomical branch is not fully covered in Sunpy. To mitigate the issue, we propose a Python package to streamline the handling of data from RATAN-600 radio astronomical observations. The RATAN-600 archive~\citep{tokhchukova2011ratan}, with data spanning over two decades, offers researchers a unique opportunity to study the Sun's behavior across multiple solar cycles. Since 1997, RATAN-600 has amassed an archive covering solar cycles 23, 24, and the ongoing 25th cycle, comprising over 30,000 observations. These 1D scans, provided in both left and right circular polarizations, span a frequency range of 3–18 GHz \citep{shendrik2020spatial}. The sheer scale of this archive represents an invaluable resource for solar physicists, yet without the proper tools to analyze this data efficiently, its full potential remains untapped.

In this paper, we present the~\texttt{RatanSunPy} library, which aims to provide a user-friendly environment for accessing and processing solar radio observations from RATAN-600. The aim of the library is to facilitate the analysis of vast dataset collected by RATAN-600. It allows for calibration of the samples in order to facilitate further analysis and comparison with other sources. The paper is organized as follows. Section 2 presents a brief overview of the RATAN-600 instrument and the solar observations obtained from it. Section 3 provides an overview of \texttt{RatanSunPy}, its main functionalities, and package structure. Section 4 discusses RATAN-600 data processing methodology, with a usage example presented in Section 5. Finally, we evaluate the quality and efficiency of the code in Section 6, provide comparisons with similar projects in Section 7 before concluding with a discussion in Section 8.

\section{RATAN-600 Solar observations}

\subsection{RATAN-600: the World's biggest reflector}
The RATAN-600 is the multi-element-reflector radio telescope composed of free-standing surface elements ~\citep{bogod2011ratan}. The antenna of the telescope is a circular structure with a diameter of 576 meters and comprises four independent sectors: Northern, Southern, Western, and Eastern. Inside the RATAN ring, there is a Flat reflector, measuring 400 meters in length and 8 meters in width. This reflector is composed of 124 individual elements, $3.1 \times 8.5$ meters each. The flat reflector allows for the creation of a periscope system to observe astronomical objects (and radio sources) in the southern part of the sky (for example, the Sun) from the South sector. The flat reflector is adjusted depending on the elevation of the observed object. After that, South sector observes the horizontal beams.~\citep{tokhchukova2014computation}. Three of these sectors can be utilized simultaneously for three different observational programs. Movable cabins equipped with secondary mirrors and receiving apparatus are positioned at the focal point of the antenna. Consequently, the radio telescope operates on a three-mirror principle: initially, a flat reflector captures the radiation and reflects it to the South antenna sector, which subsequently directs all the rays into the cabin. For regular solar observations on the RATAN-600, the Southern sector with the Flat reflector is employed \citep{parijskij1993ratan,akhmedov1982measurement, tokhchukova2014computation}.


Observations with the RATAN-600 began in 1974 and continue daily to this day. Regular, continuous solar observations are available at least from 1997, providing an extensive and  archive of historical observations that is extremely valuable for researchers. Currently, the main device  is the 80-channel spectral radiometric complex \citep{bogod2011system}, with operating frequencies ranging from 3 to 18 GHz in both right-hand (R) and left-hand (L) circular polarizations.
All available observations since 1997 can be found at the address of the Prognostic Center of the St. Petersburg branch of the Special Astrophysical Observatory of the Russian Academy of Sciences \footnote{SAO Prognostic Center: \url{http://spbf.sao.ru/prognoz}} \citep{tokhchukova2011ratan}. They are stored in the intensity (Stokes parameter $I$) and circular polarization (Stokes parameter $V$) channels \citep{yasnov2011polarized}.

\begin{equation}
I = \frac{R + L}{2},\,\,\, V = \frac{R - L}{2}
\end{equation}
A newly designed, additional spectrometer for solar observations at the RATAN-600 radio telescope has been recently developed and implemented. This advanced decimeter-band system provides comprehensive frequency coverage across the 1–3 GHz range and is capable of acquiring data with high spectral resolution through its 1024 spectral channels. The system represents a significant enhancement in observational capabilities for solar studies ~\citep{ripak2023rfi}.

\begin{figure}[ht]
\centering
\includegraphics[width=1\linewidth]{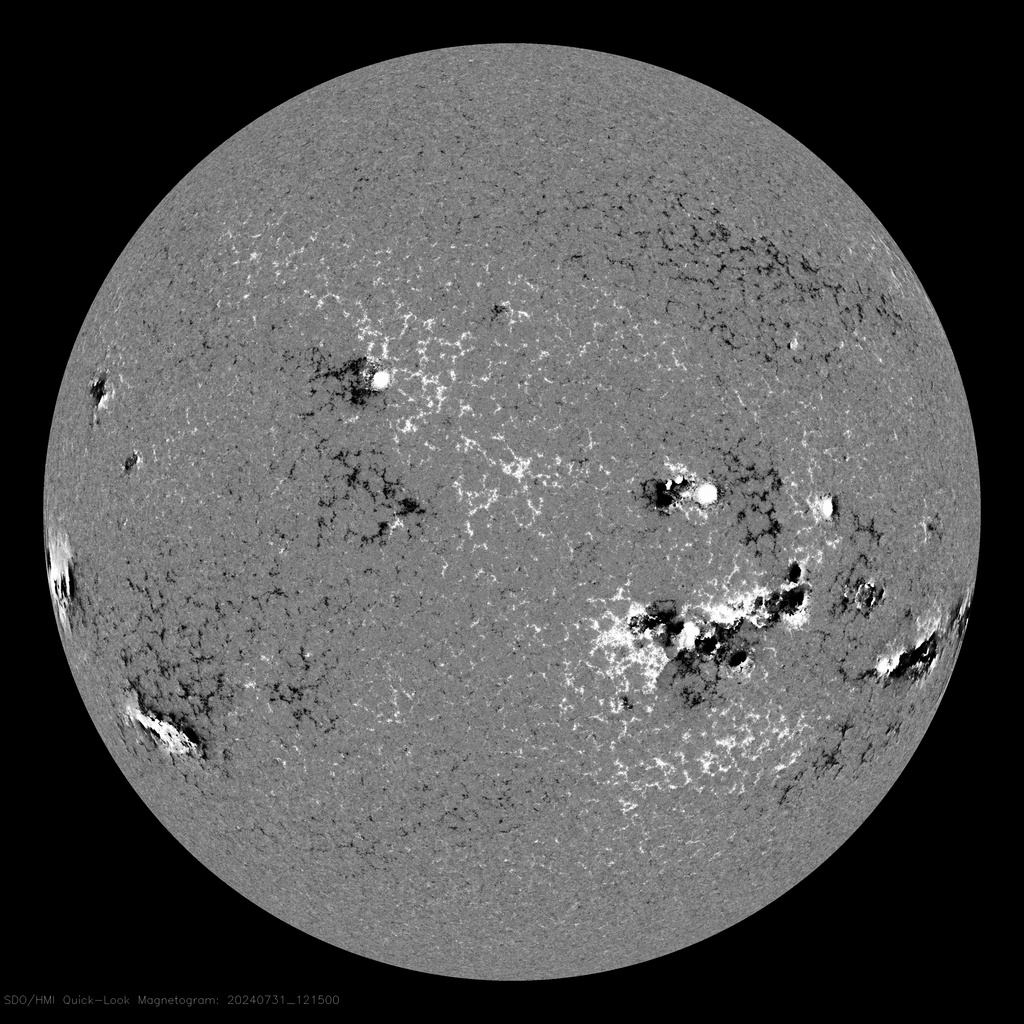}
\caption{SDO HMI magnetogram for 2024/07/31. Image was taken from  \href{SDO observatory site}{https://sdo.gsfc.nasa.gov/assets/img/latest/}}
\label{fig.SDO}
\end{figure}

\subsection{Microwave observations of the solar corona}

Solar active regions are complex three-dimensional structures encompassing the entirety of magnetic phenomena from the photosphere~\citep{nita2011threedstructure}, extending through the chromosphere, and into the corona, with various configurations of closed and open magnetic fields. These regions are associated with the processes of energy transfer, accumulation, and release, which are connected with solar flares and coronal mass ejections. The magnetic field plays a dominant role in all these processes. Direct measurements of magnetic fields at the photospheric level have been available for a long time through numerous instruments, both ground-based and space observatories. However, observations of the magnetic field structure in the chromosphere and corona remain a challenging task yet to be fully addressed~\citep{stupishin2018modeling}. Nevertheless, valuable information on coronal magnetic fields can be obtained through radio astronomical measurements of spectral polarization in the microwave range~\citep{bogod2017study}.

\begin{figure}[ht]
\centering
\includegraphics[width=1\linewidth]{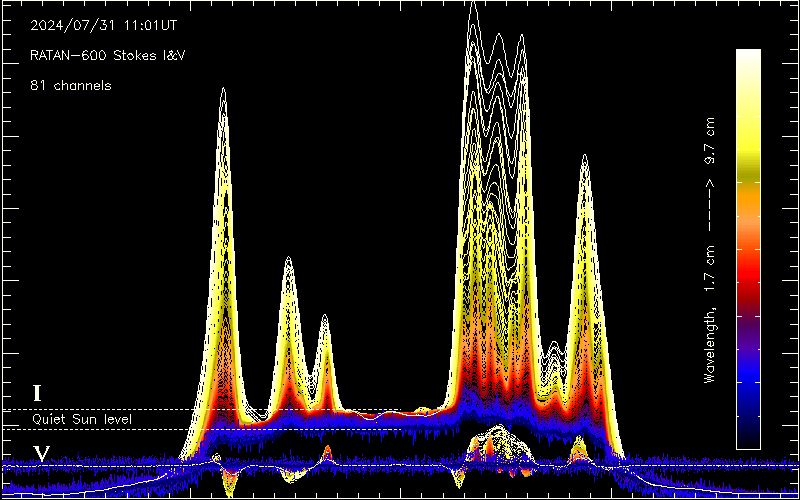}
\caption{RATAN-600 solar data for 2024/31/07, same date as for Fig. ~\ref{fig.SDO}}
\label{fig.Ratan}
\end{figure}

Photospheric data, often referred to as magnetograms, provide a two-dimensional visual representation of the Sun, as illustrated in Fig.~\ref{fig.SDO}. In these images, dark and bright regions correspond to solar active regions, enabling effective localization and tracking of their evolution over time. In contrast, solar radio data obtained from the RATAN-600 telescope are presented as one-dimensional dual-polarization, multi-wavelength scans. These scans cover a broad frequency range of 3–18 GHz, offering high spectral resolution (up to 100 MHz), spatial resolution (up to 18 arcseconds)~\citep{tokhchukova2014computation}, and remarkable sensitivity in terms of flux density, reaching values as low as 0.1 sfu~\citep{shendrik2020spatial}.  Therefore, to describe a particular active region using radio data, it is necessary to isolate the source of this region from the entire spectrum. This task must be addressed separately, for example, through Gaussian analysis. An example is provided, showing a fragment of a solar magnetogram and the corresponding radio spectrum at a specific frequency for that moment in time, see Fig.˜\ref{fig.Ratan}.

\begin{figure}[ht]
\centering
\includegraphics[width=1\linewidth]{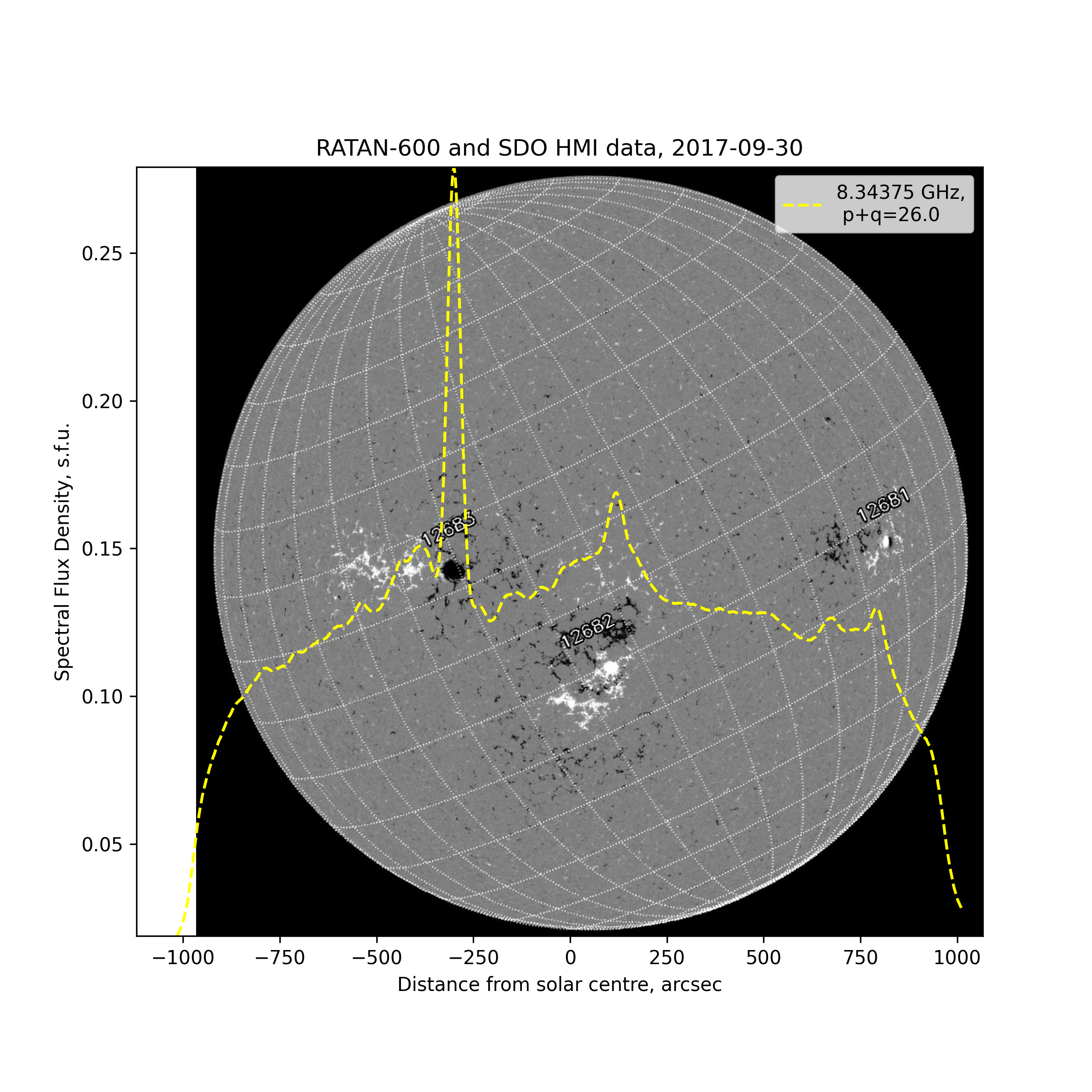}
\caption{Overlay of RATAN-600 Observation on SDO HMI Magnetogram with NOAA Active Regions Marked ~\ref{fig.SDO}}
\label{fig.RatanSDO}
\end{figure}

Fig.~\ref{fig.RatanSDO} illustrates an overlay of observations from the RATAN-600 radio telescope on a magnetogram captured by the Solar Dynamics Observatory's Helioseismic and Magnetic Imager (SDO HMI). This composite image highlights the spatial relationship between solar radio emissions detected by RATAN-600 and the magnetic field structures observed by SDO.

\section{\texttt{RatanSunPy} overview}
\label{package overview}
The Python library for working with RATAN-600 data primarily replicates the functionality available through the user web interface for interactive search, visualization, and online analysis of RATAN-600 data, which can be accessed at \href{SAO system: Prognoz}{http://www.spbf.sao.ru/prognoz/]}.

The primary data source for the module is the polarised spectra recorded using the main complex of the Sun with RATAN-600, covering the frequency range of 3–18~GHz in right and left circular polarizations. Currently, the library does not support the processing of decimeter complex data with frequencies below 3~GHz.

The observation method involves the passage of the solar disk through a stationary "knife-edge" antenna pattern. Observations are conducted simultaneously across all frequency channels, with observation time determined by the Earth's rotation speed. During the process, the signal from an empty sky is also measured, and raw observations are initially calibrated using a noise generator. Upon completion of each observation, the secondary mirror is repositioned to a new azimuth, and the process is repeated.

Initial data quality control procedures are performed on the servers of the Special Astrophysical Observatory (SAO), while the ~\texttt{RatanSunPy} module is responsible for subsequent data processing. Currently implemented functionality includes:
\begin{itemize}
    \item Data calibration, subtraction of the quiet Sun level \cite{borovik1997disser}.
    \item Automatic detection of local sources, parameters of sources estimation
    \item Identification of local sources with NOAA active regions.
    \item Calculation of active regions features in microwave radiation \cite{opeikina2015revisiting}.
\end{itemize}

\begin{figure*}[ht]
\centering
\includegraphics[width=1\linewidth]{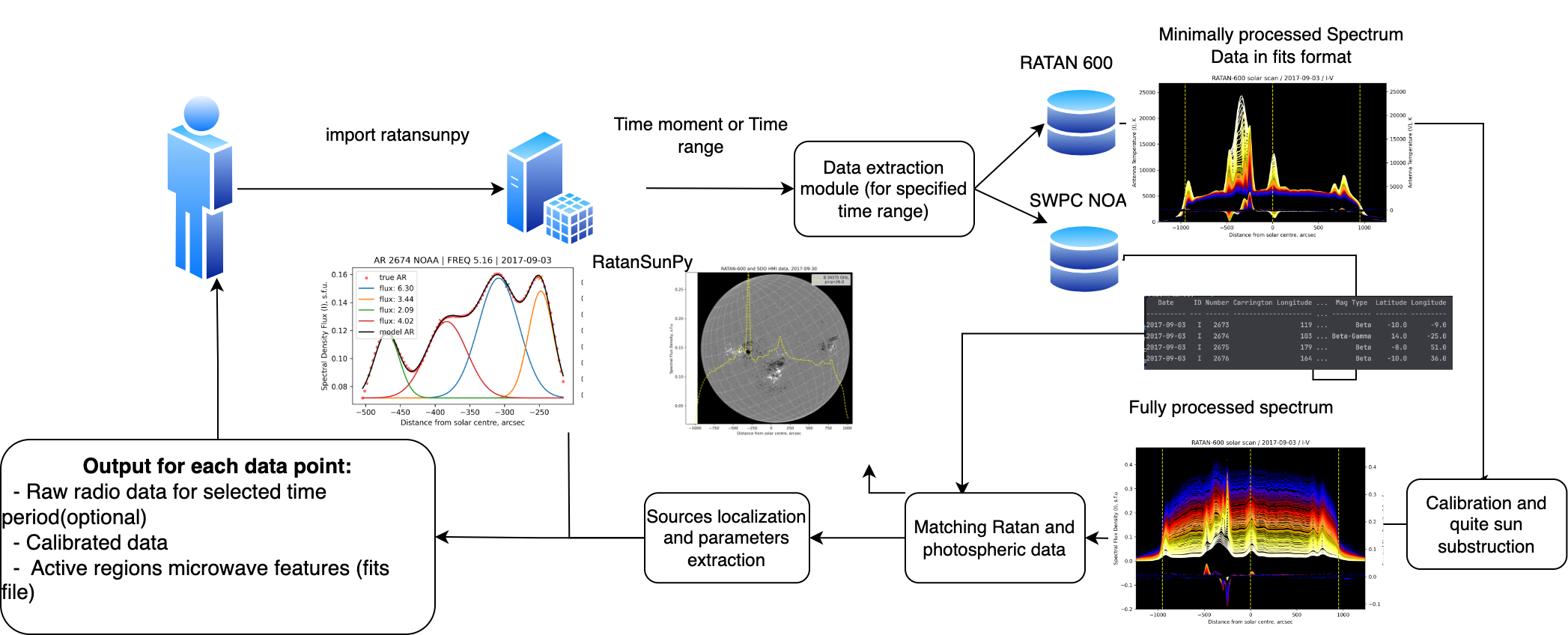}
\caption{Workflow of \texttt{RatanSunPy} package}
\label{fig.PackageInfo}
\end{figure*}

The output of the module includes processed data from the initial observation scans and tables containing the main characteristics of the local sources and active regions. The workflow of the tool is illustrated in Figure. \ref{fig.PackageInfo}. The specifics of the functionality implementation and the methodology for working with RATAN-600 data are described in the following chapter.

From a developer's perspective, the library is organized in a manner similar to the popular SunPy library, which is also written in Python and designed to work with a wide range of solar data sources. A separate client is implemented for each data source, which, using information about the time period, retrieves available data for that specified interval. 

Currently, our library includes two clients: SRSClient, designed to obtain information about active regions on the Sun from NOAA SWPC, and RATANClient, designed to work with RATAN-600 data. To search for data from a selected source over a specified time interval, a class named Scrapper is provided.

\begin{figure}[ht]
\centering
\includegraphics[width=1\linewidth]{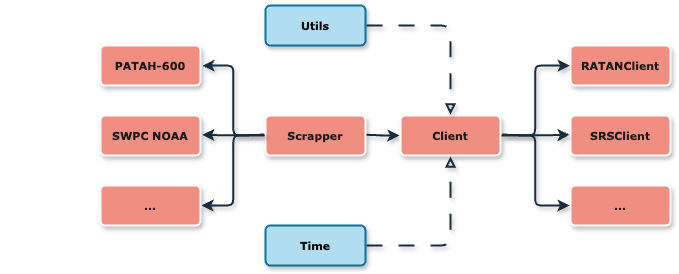}
\caption{Entity structure in \texttt{RatanSunPy}}
\label{fig.Class}
\end{figure}

To facilitate further expansion of functionality and the integration of additional data sources, an abstract class, Client, has been developed. This class serves as the base for classes responsible for working with specific instruments, such as SRSClient and RATANClient. The class organization is illustrated in Figure \ref{fig.Class}

\section{Data Processing Methodology for RATAN-600}

Utilizing radio astronomical observations from the RATAN-600 requires a systematic preparation process that involves several sequential steps. These steps include defect rejection, calibration, determining the background level, and obtaining an averaged scan of the spectral density of the microwave flux from observations of the quiet Sun. Additionally, the process involves identifying cyclotron radiation sources in solar active regions (ARs) and correlating these sources with data on ARs provided in the Solar Region Summary \footnote{\url{https://www.swpc.noaa.gov/products/solar-region-summary}}  tables from the Space Weather Prediction Center (SWPC) of the National Oceanic and Atmospheric Administration (NOAA). These preprocessing steps were originally developed at the SAO RAS  \citep{shendrik2020spatial, tokhchukova2011ratan}  and have now been reimplemented as a part of Python module in close collaboration with the core developers.

\subsection{Data Calibration}

The first and one of the most crucial stages of data processing is the calibration of the data obtained from the radio telescope. This procedure is essential because measuring the effective area of a large antenna like the RATAN-600 is often challenging or practically impossible. To obtain values close to the actual ones, it is necessary to scale the scan, which is essentially the calibration procedure (\citep{opeikina2015revisiting}).

Traditionally, RATAN-600 has employed two methods for calibrating observations obtained in units of antenna temperatures \citep{nindos1996two}:

\begin{itemize}
    \item Scaling observations from zero to the antenna temperature level of the quiet Sun at a point located in a relatively quiet part of the Sun near the scan center. The table of quiet Sun antenna temperatures is obtained separately using additional observations of the Moon.
    \item Calculating the area under the observation curve at a given frequency, which is then equated to the total flux of the Sun obtained from other radio telescopes with small mirrors. An example of such a calibration instrument is the Nobeyama polarimeters.
\end{itemize}

However, the SAO laboratory is currently developing a calibration scheme based on templates of the quiet Sun, as described below.  For consistency with existing tools for processing data from RATAN 600, we have adopted the same calibration technique in our project. However, users are free to implement their own calibration methods if desired, providing flexibility for diverse research needs. Additionally, we plan to expand the range of calibration approaches available in future updates to accommodate broader scientific applications.

\begin{figure}[ht]
\centering
\includegraphics[width=1\linewidth]{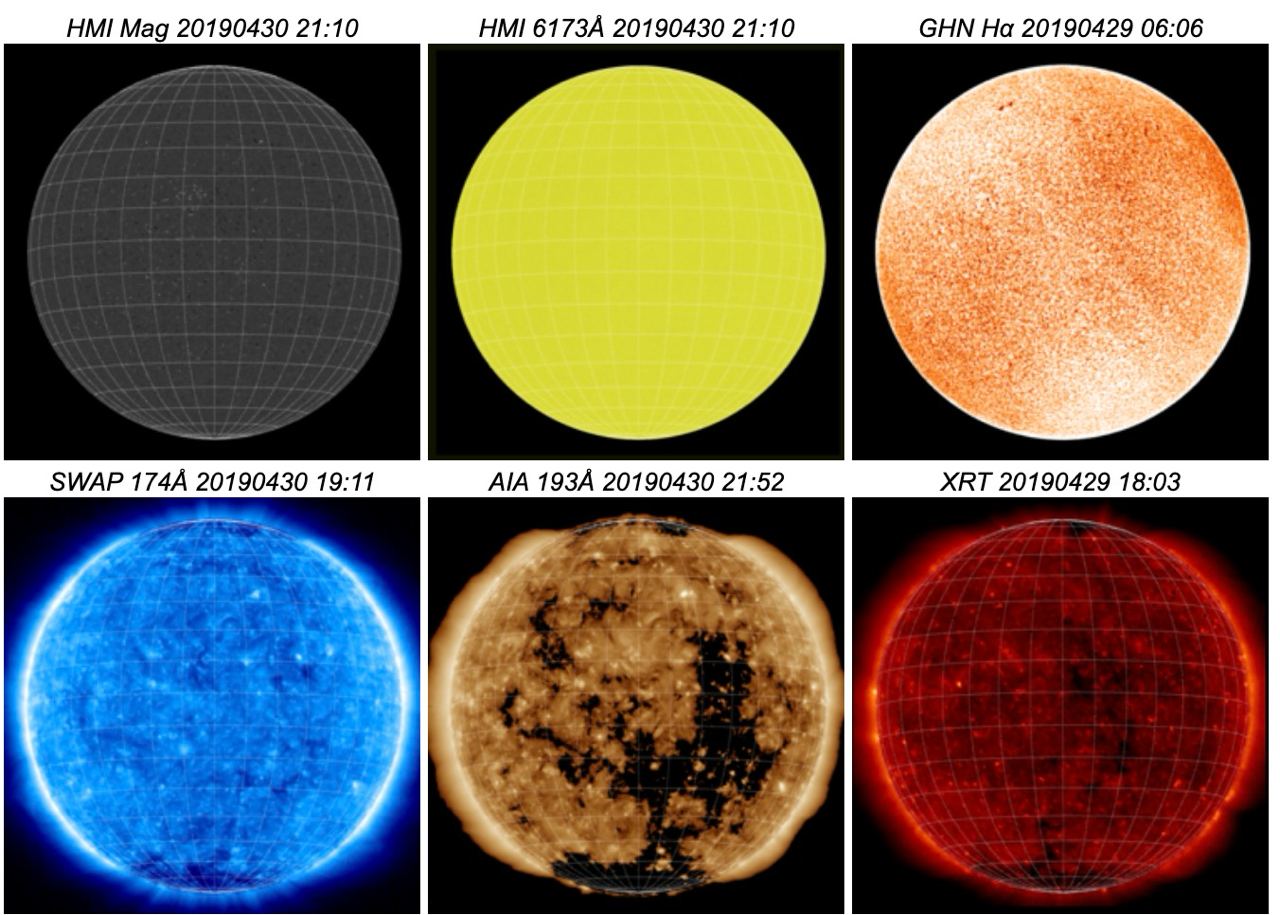}
\caption{Example of the quiet Sun, April 30, 2019}
\label{fig.QS}
\end{figure}

The idea behind the method is as follows. A series of solar observations with minimal solar activity during the minimum global solar activity cycle was selected using data from the Solar Monitor project website. An example of one such observation in various registration channels (X-ray, $H_\alpha$, ultraviolet, magnetogram) is shown in Figure \ref{fig.QS}. For each selected observation, the following steps were performed:

\begin{itemize}
\item The observations were cleaned of defective and noisy frequencies.
\item Centering was performed.
\item Calibration was done by convolving the antenna's directivity pattern with a "quiet Sun" model in the form of a disk of constant intensity.
\end{itemize}

\begin{figure}[ht]
\centering
\includegraphics[width=1\linewidth]{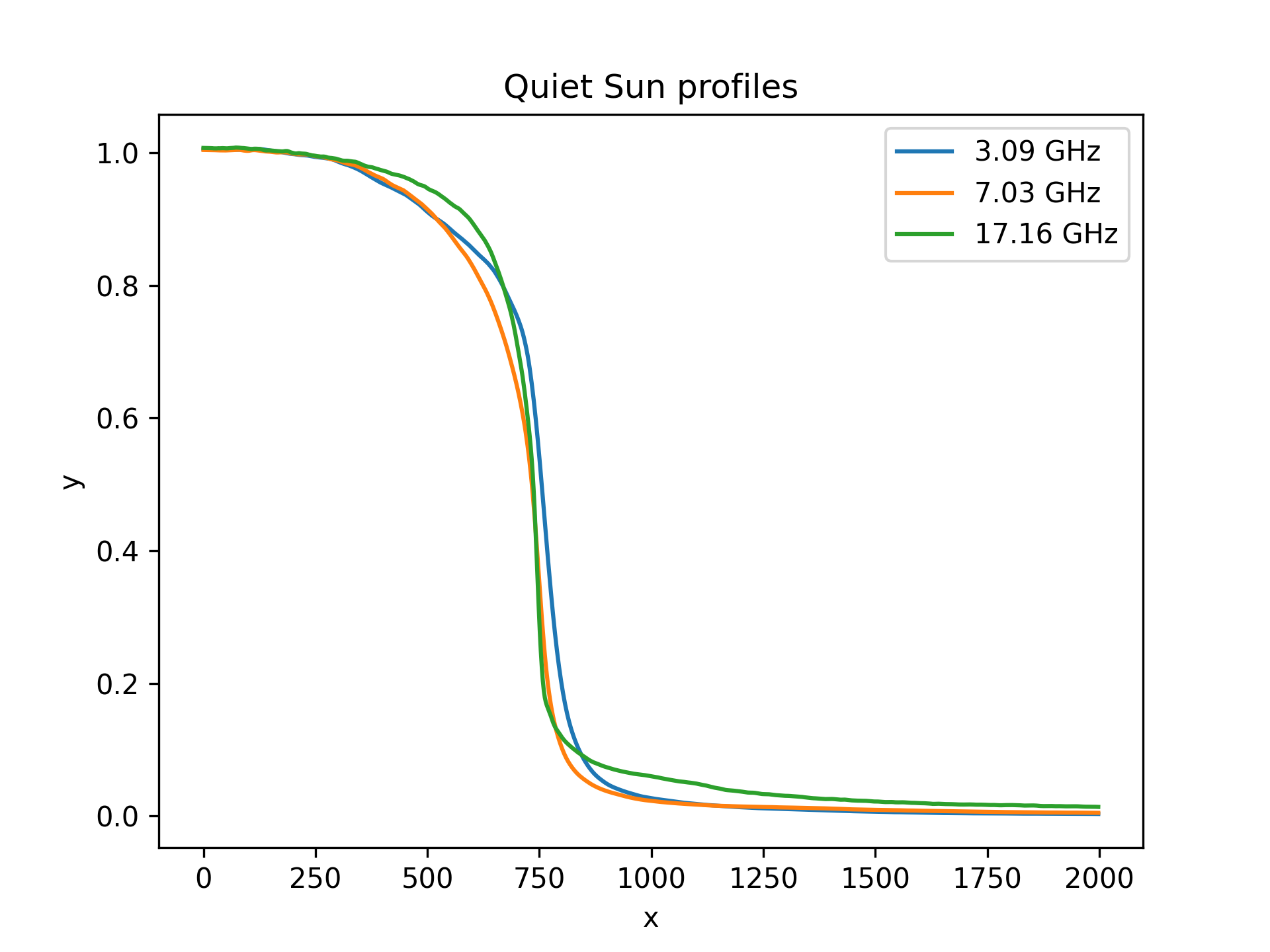}
\caption{Obtained profiles of the quiet Sun}
\label{fig.QSprofiles}
\end{figure}

All these observations were aligned to the same angular radius and a common abscissa. The mean angular radius of the solar disk is approximately 950 arcseconds. While the full solar disk fits within +-1000 arcseconds, calibration purposes require observations to extend beyond the solar disk to include regions with negligible coronal signal. A region encompassing twice the solar radius, or approximately 2000 arcseconds, is sufficient to meet these requirements. That's why the data were then limited to an interval of $[-2000, 2000]$ arcseconds from the center of the Sun and smoothed at each frequency. Finally, semi-profiles of the solar disk were formed from the resulting profiles. Values were taken as the minimum when moving from the center to the edges, as shown in Figure \ref{fig.QSprofiles}, where the x-axis represents the distance from the center of the Sun in arcseconds, and the y-axis represents the specific flux density in solar flux units (s.f.u.).

The main idea of the calibration method is to use these semi-profiles of the quiet Sun to scale "raw" radio astronomical observations. The automatic processing procedure for scans will be presented below.

\begin{figure}[ht]
\centering
\includegraphics[width=1\linewidth]{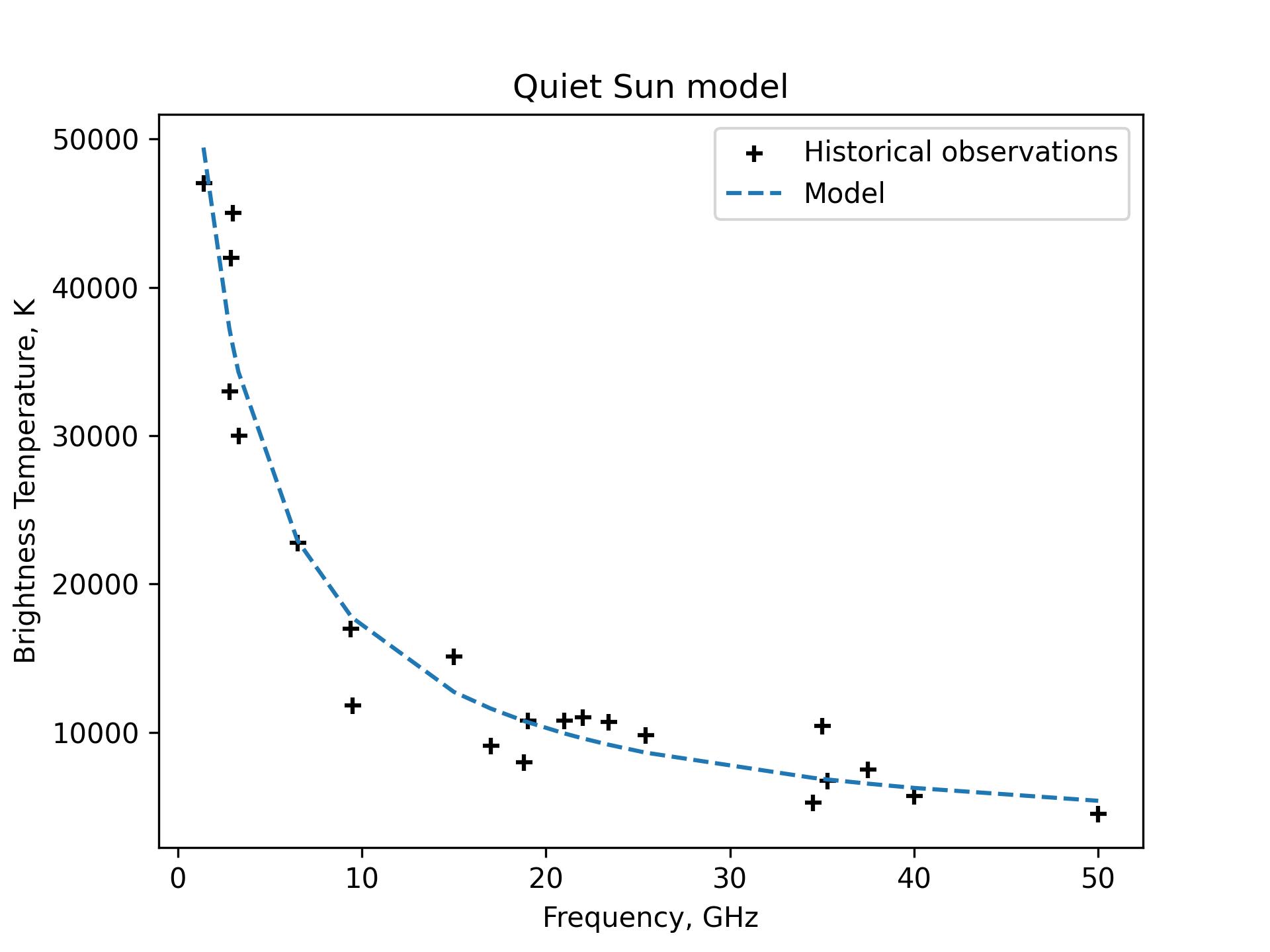}
\label{fig.qsModel}
\caption{Model of the dependence of the quiet Sun brightness temperatures on frequency based on historical data}
\end{figure}

In case of frequency mismatches in the observations, the quiet Sun data are interpolated to the nearest scan frequencies. Additionally, the templates are normalized so that the area under the curve matches the model of the quiet Sun. This model is based on the historical observations of the Sun in terms of brightness temperatures at different frequencies, as exemplified in the table presented in \cite{shendrik2020spatial}, see \ref{fig.qsModel}.

However, the templates are described in terms of solar flux units, so a transition from a brightness temperature-based model to a radio flux density model is required. This transition can be made according to the formula derived from the Rayleigh-Jeans law, which describes the spectral radiance:

\begin{equation}
F_s = B_{\nu} \Omega_s = 2 \cdot 10^{22} \cdot \frac{k_B T_b \Omega_s c^2}{\nu^2}
\end{equation}

Here, $\Omega_s$ is the solid angle of the Sun [sr], $k_B$ is the Boltzmann constant [J/K], $T_b$ is the brightness temperature [K], $c$ is the speed of light in vacuum [m/s], $\nu$ is the frequency [Hz], $B_{\nu}$ is the specific intensity of radiation [W/sr m² Hz], and $F_s$ is the spectral flux density [s.f.u.].

For a small angle, the solid angle can be expressed in degrees as follows (where $\alpha = 0.25^\circ$):

\begin{equation}
\Omega_s = \pi \left(\frac{\alpha \pi}{180}\right)^2
\end{equation}

Thus, we obtain the values of the total solar flux depending on the observation frequency and bring the areas under the template profile curves to these values.

\begin{figure}[ht]
\centering
\includegraphics[width=1\linewidth]{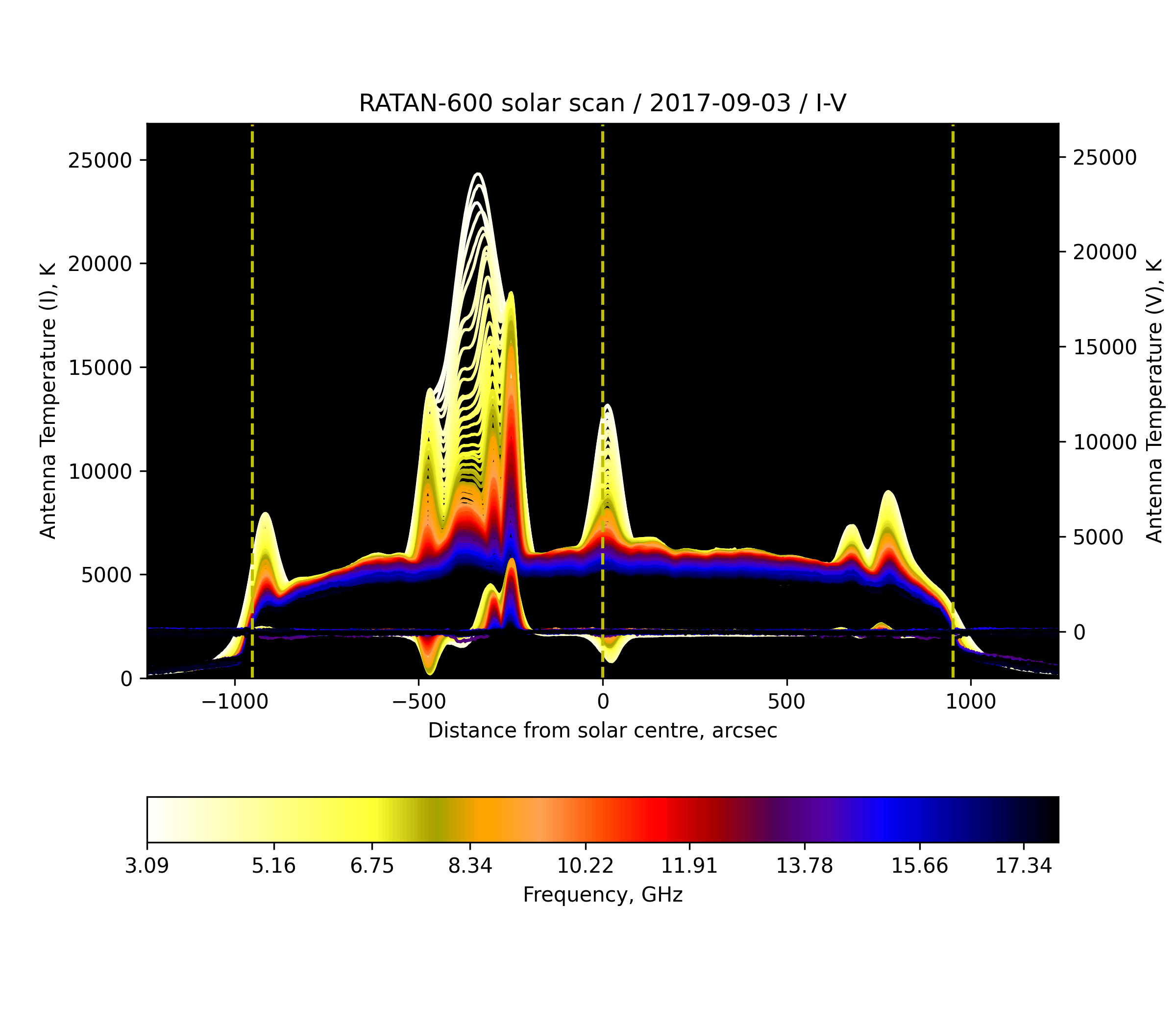}
\caption{Example of raw scan from Ratan-600}
\label{fig.RawScan}
\end{figure}

\begin{figure}[ht]
\centering
\includegraphics[width=1\linewidth]{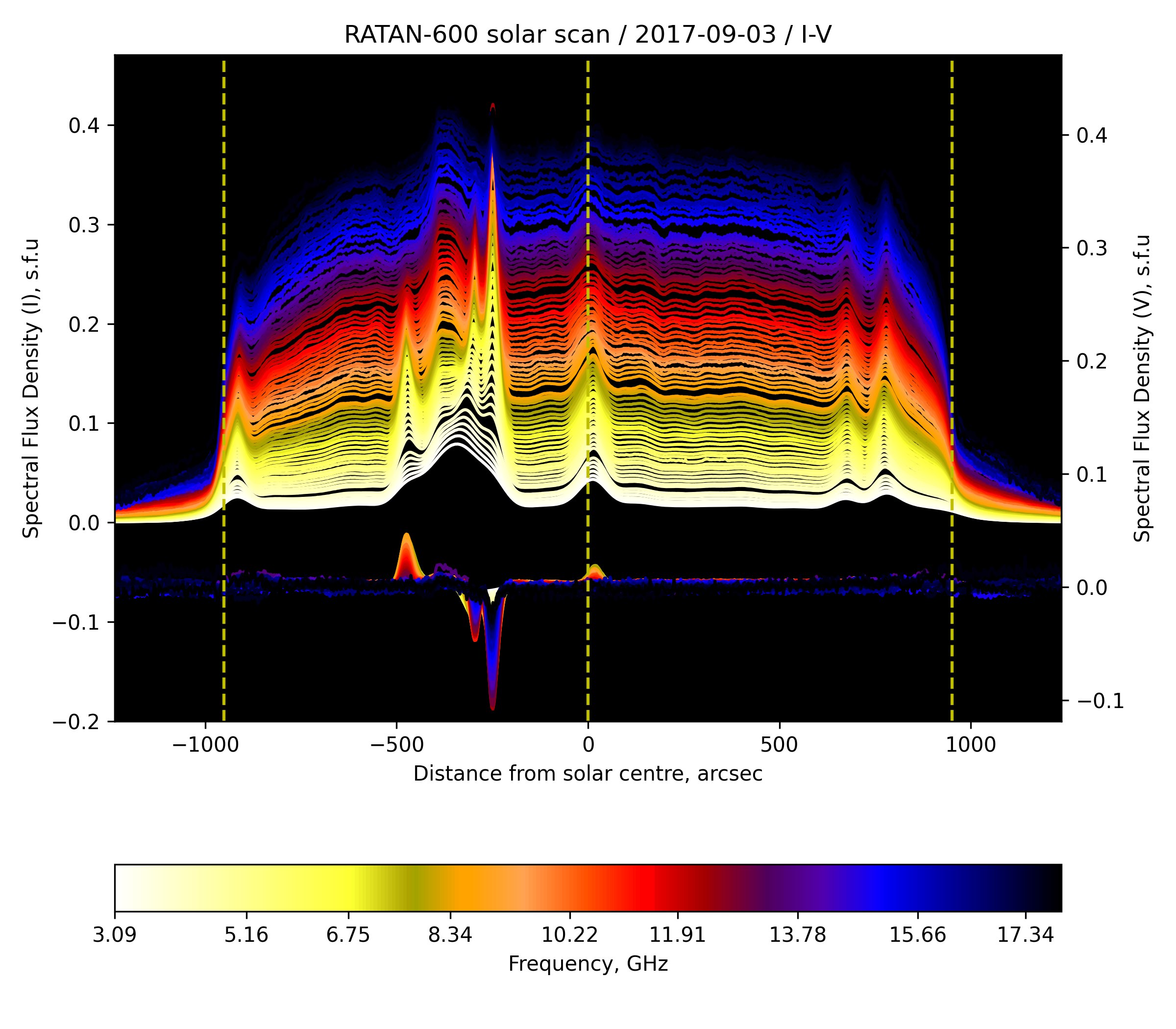}
\caption{Example of calibrated scan from Ratan-600 }
\label{fig.CalibrScan}
\end{figure}

The observation must then be scaled so that the background constant level of solar activity matches the quiet Sun template as closely as possible at a given frequency. In other words, the observation is "fitted" to the template, completing the calibration process. An example of a raw scan and the resulting spectrum after calibration, obtained using the approach described above and implemented in our library, is presented in Figure \ref{fig.RawScan} and Figure \ref{fig.CalibrScan}

\subsection{Automatic detection of local sources}
\begin{figure}[ht] 
\centering 
\includegraphics[width=1\linewidth]{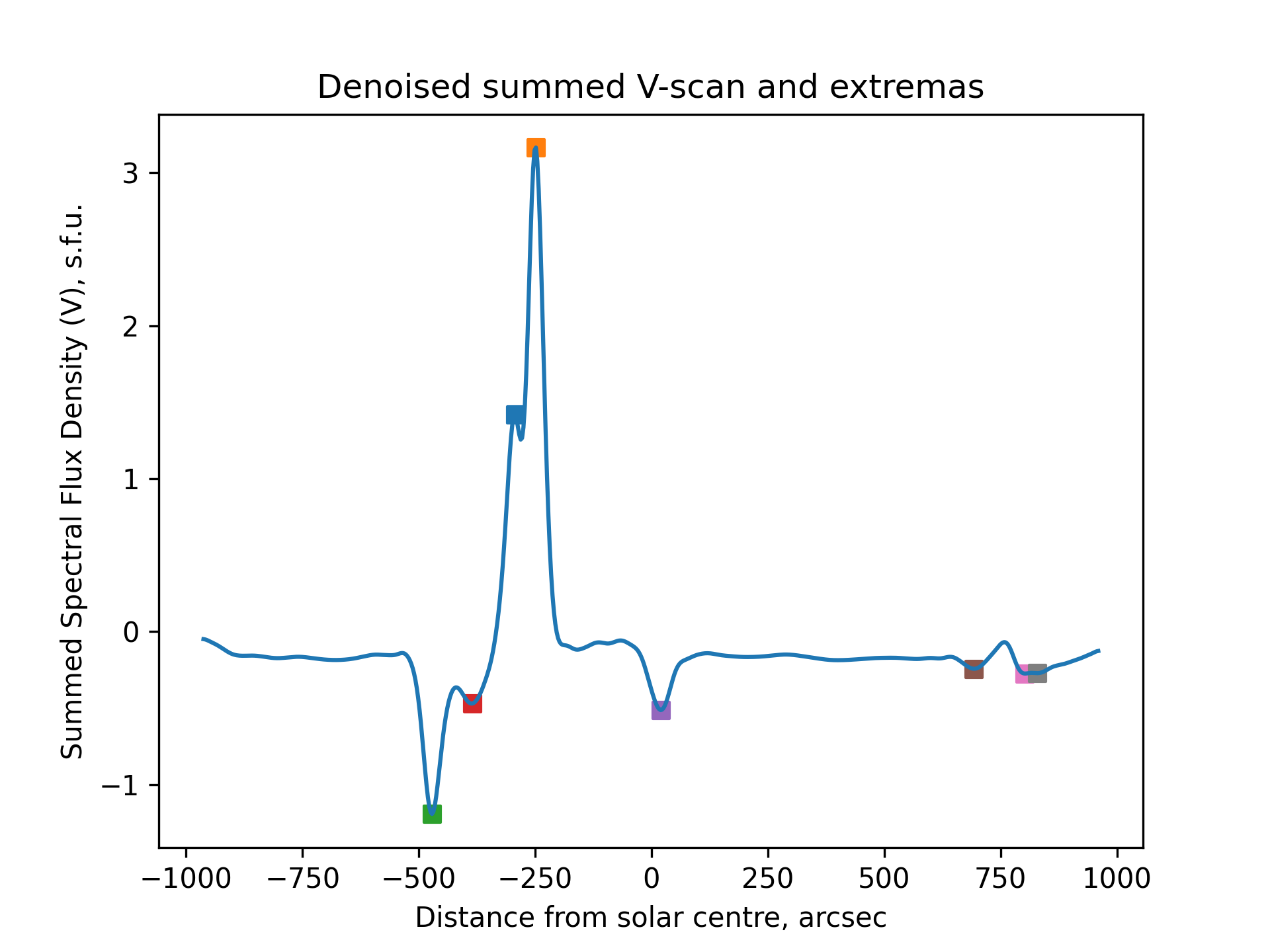} 
\caption{Denoised signal} \label{fig.Denoise} 
\end{figure}

A local source (LS) on the Sun corresponds to a local enhancement of the magnetic field and is visually characterized by a peak on the intensity graph. For automatic peak detection, the find\_peaks function from the scipy.signal library was used, which returns a set of local extrema from a one-dimensional signal. The signal used for this purpose was the data on circular polarization $V$, averaged across all frequencies. This quantity approaches zero in regions of quiet solar activity, which simplifies the task of automated peak detection. Averaging makes the peaks more pronounced. An example with the averaged signal and detected peaks is shown in ~\ref{fig.Denoise}.

However, this method has its limitations. In addition to cyclotron sources detected using their polarization differences, there are also sources with predominantly thermal emission, which cannot be localized using this approach.  Further, to improve the signal-to-noise ratio before peak detection, discrete wavelet transform was employed, specifically using Symlet wavelets. The symlets are nearly symmetrical wavelets proposed by Daubechies as modifications to the db family \citep{daubechies1988orthonormal}.

\subsection{Identification of local sources with NOAA active regions}

To identify local sources of radiation (e.g., interspot sources, sources above the leading sunspot, sources at the base of loops, etc.), it is necessary to correlate the data with observations from other observatories across different wavelength ranges. Most commonly, RATAN-600 data are compared with photospheric magnetograms (SOHO, SDO), as well as images in the ultraviolet (SOHO EIT) and X-ray (Hinode XRT) ranges.

For the alignment of one-dimensional RATAN-600 scans with two-dimensional images, it is essential to determine the position of the active region at the time closest to the RATAN-600 observations. This process involves rotating the image to match the solar position angle, aligning the center and edges of the solar disk, and, if the observation times do not match, adjusting the two-dimensional image to the correct time while accounting for the Sun’s differential rotation.

In this work, the identification of local sources is implemented using the Solar Region Summary data from the Space Weather Prediction Center (SWPC). Since the SWPC data are provided for the end of the day (2400 UTC), it is necessary to rotate the two-dimensional coordinate system to the time of the RATAN-600 observations. Additionally, a correction is required for the position angle $-(p+q)$, where $p$ is the inclination of the solar rotation axis relative to the declination circle (stored under the SOLAR\_P header in the '.fits' files) and $q$ is an angle dependent on the observation azimuth (AZIMUTH parameter) and the source declination (SOL\_DEC parameter).

Subsequently, a transformation from the heliographic to the heliocentric coordinate system is necessary. This requires knowledge of the heliographic latitude of the central point of the solar disk and the solar radius, both of which are available in RATAN-600 observations in the FITS files under the headers SOLAR\_B and SOLAR\_R, respectively. Once these steps are completed, the local sources can be successfully matched with the nearest active regions (ARs).

\subsection{Calculation of active regions features in microwave radiation.}

To compute the physical parameters of LS, it is necessary to fit a Gaussian to the source data at each frequency. Subsequently, using the parameters of these Gaussians, one can estimate physical characteristics such as flux area, maximum intensity amplitude, and others. In addition, it is possible to calculate the total flux of an AR by summing up the calculated fluxes from local sources, or alternatively, by integrating a solar flux curve within the boundaries of AR \citep{opeikina2015revisiting}.

\begin{itemize}
\item Quantity of local sources in AR: To determine the number of local sources, one can compare polarization configurations (Stokes V) or use other data, such as photospheric magnetograms.
\item Methods for detecting local sources: At least two methods, as well as their combinations, are used: Gaussian fitting and local maxima detection.
\item Flux ratio between fluxes of local sources: One important consideration is that the total flux of local sources should be equal to the total flux of the AR. The fluxes from local sources should have a physical interpretation. For example, most of the flux may be found in the halo of the AR. Typically, the halo surrounding an active region is not particularly bright; instead, it appears as a diffuse and extended source. Despite its low brightness, the halo can contribute significantly to the total flux due to its large spatial extent, thus, the contribution from the halo should not be negligible.
\item Calculating parameters: To calculate parameters such as width and brightness temperature, it is necessary to know the radiation beam pattern of the radiotelescope. However, solar observations with RATAN-600 use a complex system with three mirrors (the South sector, Flat reflector, and equipment cabin). While the beam pattern is well understood, it may vary depending on the methods of calculation or the primary feeds used (~\citep{tokhchukova2014computation}).
\end{itemize}

After localizing the sources and associating them with active regions, the physical parameters of the local sources and the entire active region are determined using a procedure known as Gaussian analysis. This procedure is based on the fact that the one-dimensional profiles of local sources, obtained by scanning with the antenna’s beam pattern, are approximately Gaussian in shape. That is, they can be modeled by functions of the form:

\begin{equation} 
f(x) = A e^{-\frac{(x-b)^2}{2 c^2}} 
\end{equation}

\begin{figure}[ht] 
\centering 
\includegraphics[width=1\linewidth]{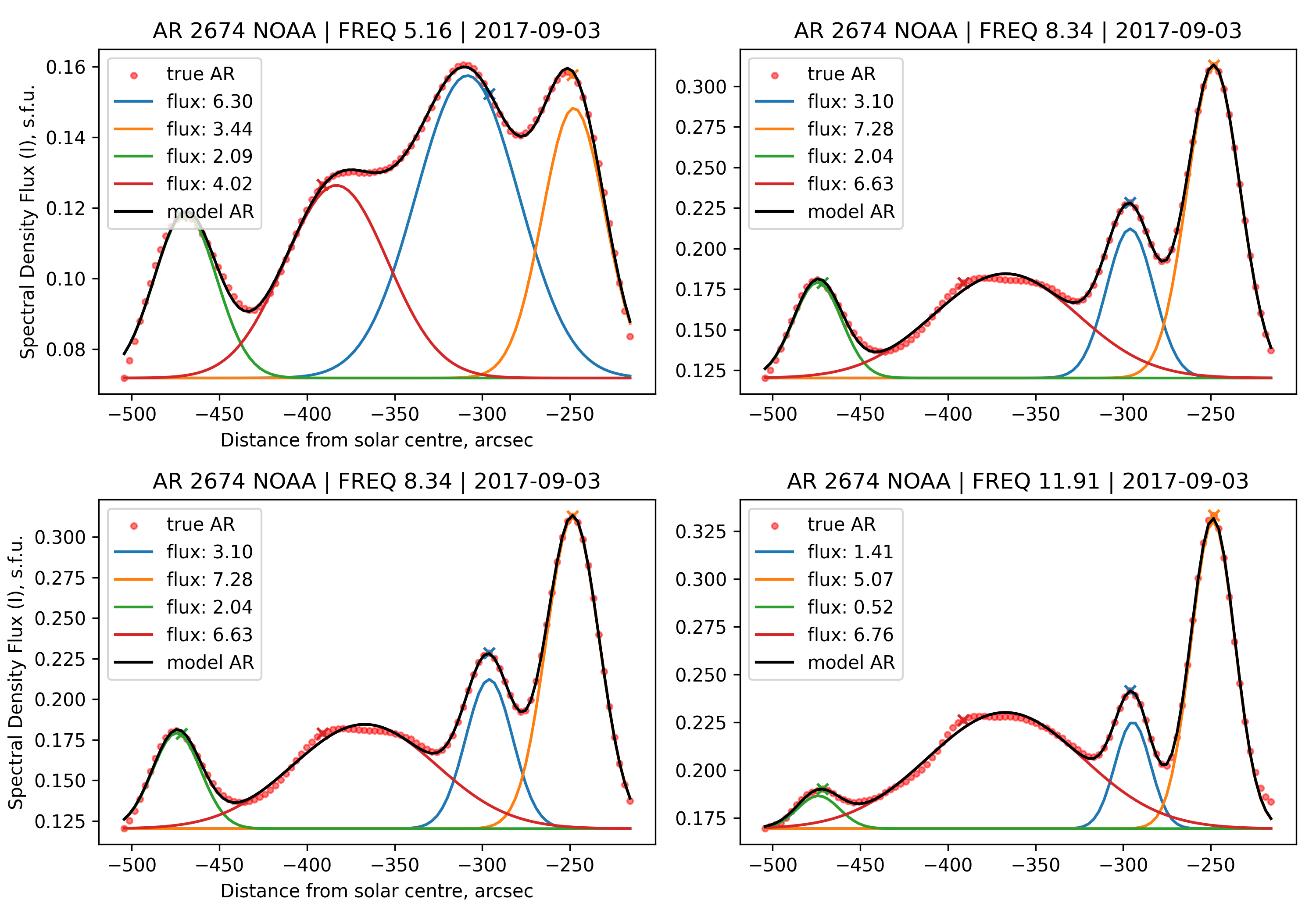} 
\caption{Gaussian analysis for AR 2674 on September 3, 2017} \label{fig.GaussWide} 
\end{figure}

The Gaussian analysis involves calculating the optimal parameters of the Gaussians such that the difference between their sum and the curve segment of the specific active region on the solar disk at the given frequency is minimized using the least squares method:

\begin{equation} 
\sum_{j=0}^{len(x)}\left(AR(x_j) - \sum_{i=0}^N F_{0_i} e^{-\frac{(x_j -\mu_i)^2}{2 \sigma_i^2}}\right)^2 \rightarrow \text{min} 
\end{equation}

where $AR(x)$ is the curve describing the segment of the active region on the solar disk at the selected frequency, $F_{0_i}$ is the amplitude of the Gaussian for the $i$-th local source, $\mu_i$ is the mean of the $i$-th local source, $\sigma_i$ is the standard deviation of the $i$-th local source, and $N$ is the number of local sources found within the active region. An illustration of the algorithm’s operation is shown in Figure \ref{fig.GaussWide}.

\begin{figure}[ht] 
\centering 
\includegraphics[width=1\linewidth]{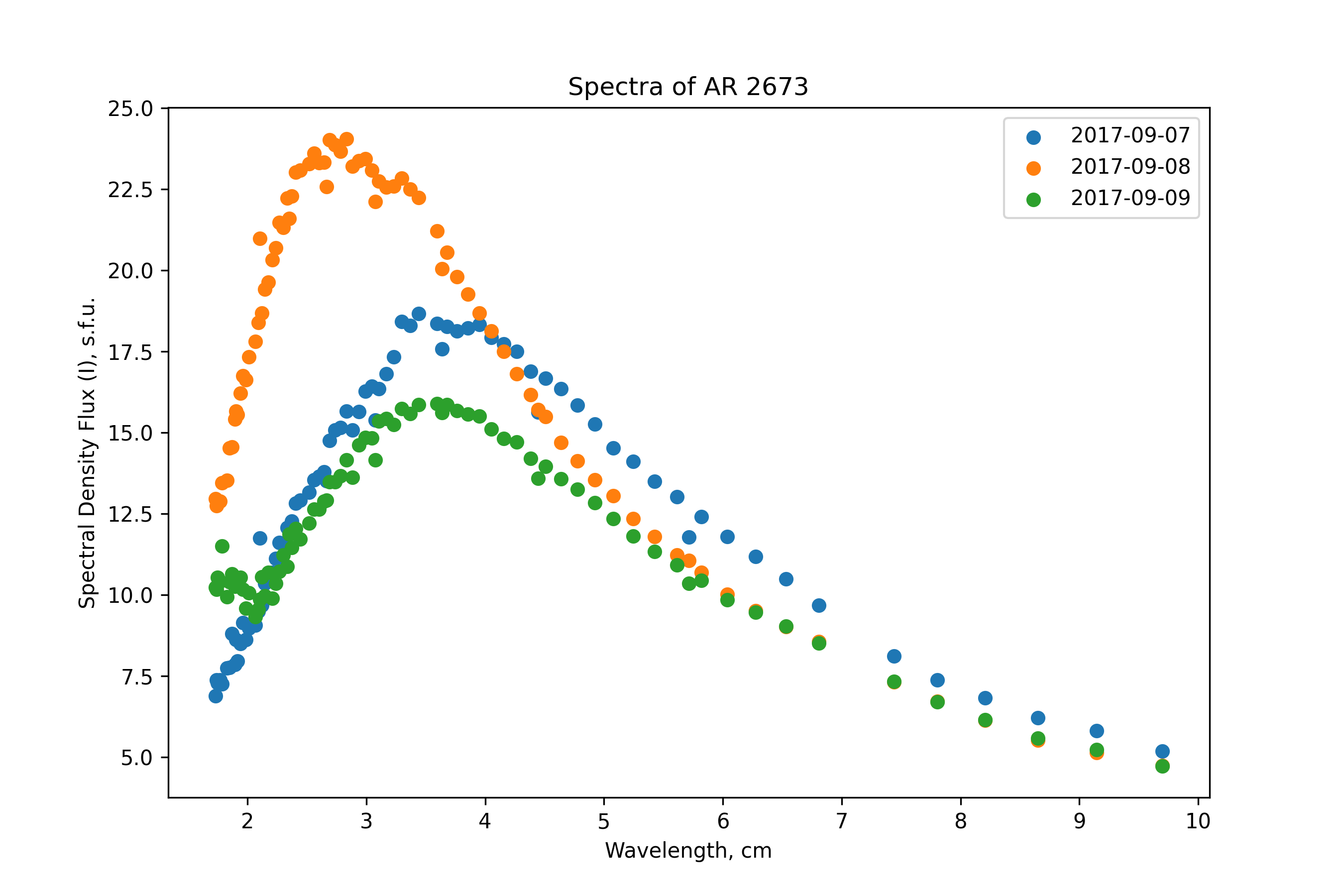} 
\caption{Three day of evolution of AR 2673} \label{fig.Spectra} 
\end{figure}

Once the Gaussian analysis is completed, it becomes possible to calculate certain physical parameters of the local sources and the associated active regions. These parameters include the total flux of the source, brightness temperatures, and source sizes. 
It is essential to study the evolution of active regions on the Sun, as these areas provide valuable insights into the Sun’s complex magnetic field dynamics and the underlying mechanisms that drive solar activity. By analyzing active regions over time, we can better understand the conditions that lead to solar flares and other solar phenomena, ultimately improving our ability to predict space weather events. Our package enables instruments for detailed comparisons and analysis of active region spectra across different wave bands, tracked over time. An example of this is shown in Figure \ref{fig.Spectra}, which illustrates the spectral evolution of Active Region (AR) 2673 over the course of three consecutive days.

\section{Usage example}
\subsection{Usage Example}
"Comprehensive usage examples can be found in the example notebooks available on the project's \texttt{github} repository \footnote{\protect\url{https://github.com/SpbfSAO/RATANSunPy}} and within the official documentation \footnote{\protect\url{https://spbfsao.github.io/RATANSunPy/}}. 

The main functionality and operations with full-disk scan includes data loading and FITS calibration. Raw data can either be stored locally or downloaded from SAO resources. An example of code for acquiring data for a specific day is shown below:

\begin{verbatim}
ratan_client = RATANClient()  
timerange = TimeRange('2017-09-03', '2017-09-03')  
url = ratan_client.acquire_data(timerange)[0]  
print(url)
\end{verbatim}

The result includes the original and calibrated scans, with all information about the scan contained in the header, and the data in the corresponding data structure. An example of processing this data is provided below:

\begin{verbatim}
raw, processed = ratan_client.process_fits_data(  
    url,  
    save_path=None,  
    save_with_original=False  
)

I = processed[1].data  
V = processed[2].data
FREQ = processed[3].data  
\end{verbatim}

Additionally, it is possible to extract information about all NOAA active regions from the NOAA SWPC Solar Region Summary reports for a specific date or timerange, as shown in the following code:

\begin{verbatim}
noaa_client = SRSClient()  
timerange = TimeRange('2017-09-03', '2017-09-04')  
url = noaa_client.acquire_data(timerange)[0]
AR_table = noaa_client.get_data(timerange)
\end{verbatim}

The following code illustrates how to handle local sources and active region information:

\begin{verbatim}
path_to_processed = '/path_to_fits.fits'  
ar_info = 
ratan_client.get_ar_info_from_processed(
                        path_to_processed)
\end{verbatim}

Here, \texttt{ar\_info} is a \texttt{BinTableHDU} object, where the data contains all information about the active regions at all available frequencies, including fluxes, amplitudes, and ranges.

\section{Quality assurance}
All classes and other key components of the code are verified through automated tests written using pytest \footnote{\protect\url{https://docs.pytest.org/en/stable/}}. These tests can be found in the repository within the "tests" directory. This method ensures that the current version of the package is free from obvious bugs. All tests in docstring were tested with doctest \footnote{\protect\url{https://docs.python.org/3/library/doctest.html}]}.


\section{Comparison with similar projects}
There are at least three established methods for processing RATAN-600 solar data: using specialized software like Workscan ~\citep{garaimov1997processing} and DataAnalyzer ~\citep{shendrik2020spatial}, or employing routines in high-level programming languages such as IDL ~\citep{landsman1993idl} and Python. These tools support a wide range of operations, including calibration, trimming, and comparison with 2D scans, offering flexibility and robust data manipulation capabilities. However, these traditional approaches often require downloading the data locally and organizing it into a specific directory structure, which can make the workflow less efficient.

The \texttt{RatanSunPy} package streamlines this process by eliminating the need for standalone software that demands familiarity with specialized interfaces. While users still need to download the data, the package provides a unified and accessible framework for processing, enabling researchers to perform all necessary analyses within a single environment. Furthermore, this Python-based infrastructure makes it possible to leverage free cloud-based platforms, such as Google Colab \footnote{\protect\url{https://colab.research.google.com/}}, for data analysis. These services allow researchers to perform computationally intensive tasks without requiring local high-performance hardware, further lowering the barrier to entry and enabling wider accessibility to RATAN-600 data processing. This integration simplifies workflows, reduces costs, and enhances the flexibility of conducting scientific research.

By leveraging Python, users gain the additional advantage of integrating RATAN-600 solar data processing with other widely-used Python packages, such as \texttt{SunPy} for solar physics analysis or other instrument developed with python for solar data handling \citep{mistryukova2023stokes}. Moreover, Python’s compatibility with machine learning and deep learning libraries, like TensorFlow \footnote{\protect{\url{https://www.tensorflow.org/}}}, PyTorch \footnote{\protect\url{https://pytorch.org/}}, and Scikit-learn \footnote{\protect\url{https://scikit-learn.org/}}, offers the opportunity to apply advanced data-driven techniques to solar data, enabling automated feature detection, pattern recognition, and predictive modeling. This seamless integration of RATAN-600 data with Python’s expansive ecosystem makes it a highly efficient and versatile tool for both traditional solar data treatment and cutting-edge research.

\section{Discussion}

This work has resulted in the development of an independent Python library designed to automate the collection, processing, and analysis of solar observation data from the RATAN-600 telescope. As a valuable resource in solar radio astronomy for over two decades, RATAN-600's vast dataset requires modern tools for efficient handling. The library enables comprehensive analysis of full-disk solar spectra across multiple frequencies, allowing researchers to detect sources, characterize their physical properties, and identify active regions.
The library’s significance is further underscored by its potential in solar physics research. It simplifies the development of automated systems for predicting solar flares and other solar phenomena. Its compatibility with other Python libraries, including those for machine learning, enables the application of advanced algorithms like neural networks, enhancing data analysis capabilities. This integration with broader Python ecosystems ensures that the tool can be easily adopted, fostering collaborations and accelerating solar research.
Some challenges remain, particularly in the areas of source localization and Gaussian analysis, where instability has led to unreliable results in certain cases. Improving the stability and reliability of these components is a priority, as they are essential for accurate solar source identification and characterization.
Looking forward, the library will expand its capabilities, with plans to introduce additional calibration methods, extend support to frequencies below 3 GHz, and integrate data from other observational instruments. This will enhance the accuracy and flexibility of the analysis, allowing for cross-instrument comparisons and a more comprehensive understanding of solar phenomena.
Future developments will also include expanding the analysis of active regions, with features to track their temporal evolution, study magnetic field dynamics, and analyze spectral properties across multiple wavelengths. These enhancements will enable better identification of active regions, which often precede solar flares, and provide deeper insights into the mechanisms driving solar activity.
In conclusion, this project lays a strong foundation for future advancements in solar data analysis. As the library evolves, it will become an valuable tool for solar physicists, contributing to both the prediction and deeper understanding of solar phenomena.
\noindent\section*{Declaration of competing interest}
The authors declare that they have no known competing financial interests or personal relationships that could have appeared
to influence the work reported in this paper.
\section*{Data availability}
The data for which this project is designed to process is publicly available. Raw data from RATAN-600 available from the site of St. Petersburg branch of the SAO  \footnote{\protect\url{http://spbf.sao.ru/data/ratan/}}. Solar Region Summary available from Space Weather Prediction Center Of National Oceanic And Atmospheric Administration via http or ftp \footnote{\protect\url{https://www.swpc.noaa.gov/content/data-access}}.
\section*{Acknowledgements}
The authors would like to thank the colleagues who participated in the development and testing of the information system \href{http://www.spbf.sao.ru/prognoz/}{Prognoz}, on the basis of which this package was implemented, as well as the senior researcher of the St. Petersburg branch of the SAO RAS Bogod V.M. for the support of this work. 
We acknowledge the essential contributions of large Python packages, such as \texttt{NumPy} \footnote{\url{https://numpy.org/}}, \texttt{SciPy} \footnote{\url{https://scipy.org/}}, \texttt{AstroPy} \footnote{\url{https://www.astropy.org/}} and \texttt{Matplotlib} \footnote{\url{https://matplotlib.org/}}, which are critical for the functionality of the RATAN-600 data processing pipeline, providing key tools for numerical analysis and visualization.

\section*{Funding}
The research was carried out under the Russian Science Foundation grant No.24-21-00476, \href{https://rscf.ru/project/24-21-00476/}{Link to the project card.}


\bibliographystyle{elsarticle-harv} 
\bibliography{ratan}




\end{document}